\documentclass[final,numberedheadings]{aipproc}

\usepackage{amsmath,amssymb}
\layoutstyle{6x9}

\begin{document}

\title{Dark Matter and Higgs Sector
\footnote{To appear in the Proceedings of the sixth International
Workshop on the Dark Side of the Universe (DSU2010) Leon,
Guanajuato, Mexico 1-6 June 2010.} }
\author{Jose A. R. Cembranos}{
  address={William I. Fine Theoretical Physics Institute,
University of Minnesota, Minneapolis, 55455, USA},
 altaddress={
School of Physics and Astronomy, University of Minnesota,
Minneapolis, 55455, USA} }
\author{Jose H. Montes de Oca Y.}
{address={Facultad de Ciencias F\'{\i}sico-Matem\'aticas,
Benem\'erita Universidad Aut\'onoma de Puebla,\\ Puebla, Pue., C.P.
72570, Mexico}}
\author{Lilian Prado}
{address={Facultad de Ciencias F\'{\i}sico-Matem\'aticas,
Benem\'erita Universidad Aut\'onoma de Puebla,\\ Puebla, Pue., C.P.
72570, Mexico}}

\begin{abstract}

The inert doublet model is an extension of the Standard Model of
Elementary Particles that is defined by the only addition of a
second Higgs doublet without couplings to quarks or leptons. This
minimal framework has been studied for many reasons. In particular,
it has been suggested that the new degrees of freedom contained in
this doublet can account for the Dark Matter of the Universe.

\end{abstract}



\maketitle

\section{Introduction}
\label{intro}

In spite of many and continuous efforts, the ultraviolet (UV)
completion of the gravitational interaction is still an open
question. In these conditions, it is difficult to make general
statements about very early cosmology (although, in general,
different types of new scalar fields are commonly predicted
\cite{stgen,sft1}). However, these details are not needed to compute
the relic density of many Dark Matter (DM) candidates
\cite{bergstrom,kolb,srednicki,gondolo}. Although there are other
possibilities \cite{other}, DM is usually assumed to be in the form
of stable Weakly-interacting massive particles (WIMPs) that
naturally freeze-out with the right thermal abundance. One of the
most interesting features of WIMPs, is that they emerge in
well-motivated particle physics scenarios as in R-parity conserving
supersymmetry (SUSY) models~\cite{SUSY1,SUSY2}, universal extra
dimensions (UED)~\cite{UED1, UED2}, or brane-worlds~\cite{BW1,BW2}.
Another interesting property of WIMPs, it is that they can be tested
with high energy experiments as the new generation of colliders
\cite{Coll}. All these interesting features are also shared by the
DM provided inside the Inert Higgs Doublet Model.

\section{Inert Higgs Doublet Model (IHDM)}
The two-Higgs-doublet model (2HDM) is one of the simplest extensions
of the Higgs mechanism of the electroweak symmetry breaking beyond
the standard model (SM) \cite{lee,glashow,donoghue,haber,hall}. In
this model, an additional Higgs doublet field is introduced in the
Higgs sector of the SM with the same quantum numbers. We use the
notation
\begin{equation}
\Phi _{a}=\left(
\begin{array}{c}
\phi _{a}^{+} \\
\phi _{a}+i\chi _{a}
\end{array}
\right),   \label{doublets}
\end{equation}
where $a = 1, 2$ for the Higgs doublets before the spontaneous
symmetry breaking (SSB); that is, in the electroweak basis. Here
$\phi^{+}_{a},\phi_{a}$ and $\chi_{a}$ represent Higgs fields.
After SSB, these Higgs fields interact with the matter fields and
also self-interact via an appropriate Higgs potential. This model
presents new physics such as Flavor Changing Neutral Currents (FCNC)
and new types of CP violation
\cite{lee,glashow,donoghue,haber,hall}. Nevertheless, we are
interested in a DM candidate; so this candidate must be a neutral
scalar field which weakly interacts with other particles.

In the literature, the classification of the 2HDM is done
considering the Higgs-fermion couplings given by the Yukawa terms in
the Lagrangian \cite{kane}. For the IHDM just one Higgs doublet
interacts with the fermions in the Yukawa terms as follows
\begin{equation}
\mathcal{L}_{Y}=Y_{ij}^{(u)}\overline{q }_{Li}\tilde{\Phi
}_{1}u_{Rj}+Y_{ij}^{(d)} \overline{q }_{Li}\Phi
_{1}d_{Rj}+h.c.+\textrm{leptons},  \label{yukawa}
\end{equation}
where $\widetilde{\Phi }_{1}=i\sigma _{2}\Phi _{1}^{\ast }$,
$Y_{ij}^{(u,d)}$ are the Yukawa couplings for $i,j=1,2,3$; $u_{Ri}$,
$d_{Ri}$ are the right singlets quarks and $q_{Li}$  are left
doublets under the electroweak symmetry group \cite{kane}.

Now, for the potential we shall assume an additional $Z_{2}$
discrete symmetry\footnote{One doublet changes its sign under this
$Z_2$ symmetry.} in order to avoid the mixing term
$\Phi_{1}^{\dagger}\Phi_{2}$, which could introduce CP violation in
the potential. Then, the potential under this last condition is
\begin{eqnarray}
\mathcal{V} &=&\mu _{1}^{2}\Phi _{1}^{\dagger }\Phi _{1}+\mu
_{2}^{2}\Phi _{2}^{\dagger }\Phi _{2}+\lambda _{1}(\Phi
_{1}^{\dagger }\Phi
_{1})^{2}+\lambda _{2}(\Phi _{2}^{\dagger }\Phi _{2})^{2}  \nonumber \\
&&+\lambda _{3}(\Phi _{1}^{\dagger }\Phi _{1})(\Phi _{2}^{\dagger
}\Phi _{2})+\lambda _{4}(\Phi _{1}^{\dagger }\Phi _{2})(\Phi
_{2}^{\dagger }\Phi
_{1})  \nonumber \\
&&+\frac{1}{2}\lambda _{5}[(\Phi _{1}^{\dagger }\Phi _{2})^{2}+(\Phi
_{2}^{\dagger }\Phi _{1})^{2}],  \label{pot}
\end{eqnarray}
where $\mu_{1,2}^{2}$, $\lambda_{1,2,3,4}$ are real parameters,
while $\lambda_5$ could be complex. The $\Phi_{1}$ obtains a vacuum
expectation value (VEV): $v/\sqrt{2} = 174$ GeV, as in the SM, while
$\Phi_{2}$ does not obtain a VEV \cite{degee, kominis,dolle}. Since
this $Z_{2}$ symmetry is unbroken, the inert particles will be
stable \cite{barbi}. After SSB and by changing to the physical
basis, the Yukawa terms (\ref{yukawa}) have the form \cite{kane}:
\begin{eqnarray}
\mathcal{L}_{Hf\overline{f}} &=&-\frac{g}{2M_{W}\sin \beta }\overline{d}%
M_{d}d(H^{0}\sin \alpha +h^{0}\cos \alpha )-\frac{ig\cot \beta }{2M_{W}}%
\overline{d}M_{d}\gamma _{5}dA^{0}  \nonumber \\
&&-\frac{g}{2M_{W}\sin \beta }\overline{u}M_{u}u(H^{0}\sin \alpha
+h^{0}\cos \alpha )+\frac{ig\cot \beta
}{2M_{W}}\overline{u}M_{u}\gamma _{5}uA^{0}
\nonumber \\
&&+\frac{g\cot \beta }{2\sqrt{2}M_{W}}(H^{+}\overline{u}[M_{u}V_{CKM}(1-%
\gamma _{5})-V_{CKM}M_{d}(1-\gamma _{5})]d+h.c.  \label{Lfisico}
\end{eqnarray}
where $\alpha$ and $\beta$ angles are the mixing angles for the
Higgs bosons, $V_{CKM}$ is the CKM matrix and $H^0$, $h^0$ are
neutral CP even scalar Higgs bosons while $A^0$ is a neutral CP odd
pseudo scalar Higgs boson.

\section{Neutral Higgs as Dark Matter candidate}
\begin{figure}
  \includegraphics[scale=1]{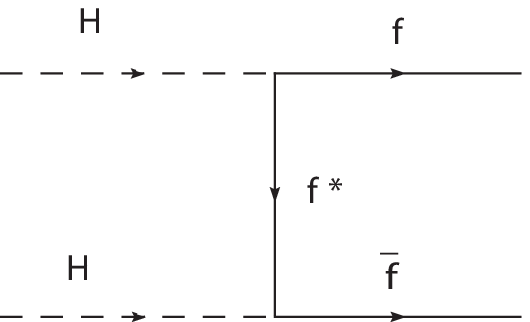}
  \hspace{0.5cm}
  \includegraphics[scale=1]{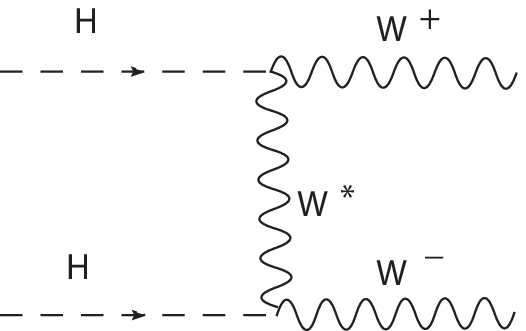}
  \caption{Feynman diagrams for the considered processes. $H$ represents the scalar or pseudoscalar neutral Higgs bosons.
  On left, the products of the annihilation are fermion and antifermion pairs. On the right, the products are $W$ charged bosons.}\label{f1}
\end{figure}
The physical fields for the Higgs sector are given by two charged
scalar Higgs bosons ($H^{\pm}$), two neutral CP even scalar Higgs
bosons ($H^0$ and $h^0$) and one neutral CP odd pseudo scalar Higgs
boson ($A^0$) \cite{kane}. Due to its characteristics of being inert
and stable, the lightest neutral scalar Higgs boson or the
pseudoscalar Higgs boson of the IHDM have been proposed as DM
candidates \cite{ma, cirelli, barbi, nezri, dolle, thomas}. Only the
scalar part $h^{0}$ has been considered for numerical results
performed using the programs MICROMEGAS and CALCHEP; the
pseudo-scalar part $A^{0}$ is only mentioned to behave in a similar
way \cite{dolle, barbi, nezri, thomas}.

\begin{figure}
  \includegraphics[scale=1]{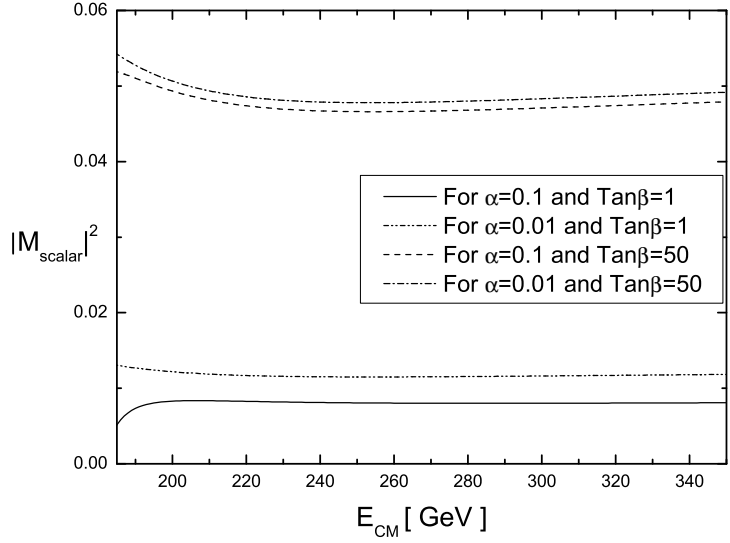}\\
  \caption{The graphic shows the numerical value for $\left\vert\mathcal{M}_{\textrm{scalar}}\right\vert^2$ as
  function of the energy in the center of mass frame. We are considering that $E_{CM} > M_{top}$. }\label{f2}
\end{figure}

In order to explore the possibilities and differences of the
pseudo-scalar part, our aim is to calculate the relic abundance for
both, $h^{0}$ and $A^{0}$, using an analytical method
\cite{srednicki,gondolo}. The relic density computation requires the
neutral Higgs bosons (of the IHDM) annihilations in standard model
pairs. We shall consider the relevant processes for the case: scalar
and pseudoscalar annihilations. In this approximation the relevant
processes are $H H\rightarrow f \bar{f}$ and $H H\rightarrow
W^+W^-$, where $H$ could be $h^0$ or $A^0$ and $f$ stands for
fermion (we just take the top quark). For example, in the limit
$E_{CM} > M_{top}$, the amplitude square for the scalar case is
given by:

\begin{eqnarray}
\left\vert\mathcal{M}_{\textrm{scalar}}\right\vert^2 &=&\left(
\frac{gM_{t}\sin \alpha }{2M_{W}\sin
\beta }\right) ^{4}\frac{\left( 6E_{CM}^{2}M_{t}^{2}-8M_{t}^{4}\right) }{%
\left( E_{CM}^{2}-M_{t}^{2}\right) ^{2}}  \notag \\
&&+\frac{\left( gM_{W}\sin \left( \beta -\alpha \right) \right)
^{4}}{\left(
E_{CM}^{2}-M_{W}^{2}\right) ^{2}}\left( 2+\left( \frac{\frac{1}{2}E_{CM}^{2}-M_{W}^{2}}{%
M_{W}^{2}}\right) ^{2}\right)   \label{Ms}
\end{eqnarray}

and for the pseudoscalar:

\begin{equation}
\left\vert\mathcal{M}_{\textrm{pseudoscalar}}\right\vert^2=\left(
\frac{gM_{t}\cot \beta }{2M_{W}} \right) ^{4}\frac{\left(
2E_{CM}^{2}M_{t}^{2}-8M_{t}^{4}\right)}{\left(
E_{CM}^{2}-M_{t}^{2}\right) ^{2}},   \label{Mp}
\end{equation}
where $E_{CM}$, $M_t$ and $M_W$ are the energy in the center of mass
frame, the top quark mass and $W$ boson mass, respectively. The
behavior of the amplitude for scalar case is shown in Figure
\ref{f2}. We have taken some characteristic values for the free
parameters $\alpha$ and $\beta$ based on \cite{maria}.

\section{Summary}

In this work we have obtained the amplitude for the neutral Higgs
bosons annihilation under a tree level approximation. The
expressions for scalar and pseudoscalar cases have a pole when the
$E_{CM}$ is equal to the top quark mass. The mixing angles are
stable for the amplitude. For the pseudoscalar case we did not
consider the second diagram contribution in order to maintain CP
symmetry in this part. However, it is necessary to consider it to
obtain a more precise value for the amplitude. Currently, we are
working in a more complete expression for amplitudes in order to
compute the discussed relic abundances.

\begin{theacknowledgments}
This work is supported in part by CONACYT, DOE grant
DOE/DE-FG02-94ER40823, FPA 2005-02327 project (DGICYT, Spain),
CAM/UCM 910309 project, and MICINN Consolider-Ingenio MULTIDARK
CSD2009-00064.
\end{theacknowledgments}

\bibliographystyle{aipproc}   

\end{document}